\documentclass[twocolumn,tighten,times]{aastex631}
\usepackage{CJK}
\usepackage[T1]{fontenc}
\usepackage{enumitem}

\newcommand\ergs{erg~s$^{-1}$}
\newcommand\ergcms{erg~cm$^{-2}$~s$^{-1}$}

\newcommand\kms{km~s$^{-1}$}

\newcommand{\hbeta}{H{$\beta$}}
\newcommand{\halpha}{H{$\alpha$}}
\newcommand{\OIII}{[O\,{\sc iii}]}

\newcommand{\OIIIb}{[O\,{\sc iii}]\,$\lambda$5007}

\newcommand{\HeII}{He\,{\sc ii}}
\newcommand{\HeIIwave}{He\,{\sc ii}\,$\lambda$4686}
\newcommand{\HeIII}{He\,{\sc iii}}

\newcommand{\NII}{[N\,{\sc ii}]}

\newcommand{\SII}{[S\,{\sc ii}]}

\newcommand{\SIIa}{[S\,{\sc ii}]\,$\lambda$6716}
\newcommand{\SIIb}{[S\,{\sc ii}]\,$\lambda$6731}

\newcommand{\HII}{H\,{\sc ii}}

\shorttitle{Photoionization of the Nebula around NGC~5408 X-1}
\shortauthors{Guo et al.}
 
\begin{document}
\begin{CJK*}{UTF8}{gbsn}

\title{Photoionization of the Composite Nebula Surrounding NGC~5408 X-1: Implications for Beamed Emission}

\author[0000-0001-9346-3677]{Jing Guo (郭静)}
\affiliation{Department of Astronomy, Tsinghua University, Beijing 100084, People's Republic of China}

\author[0000-0001-7584-6236]{Hua Feng}
\affiliation{State Key Laboratory of Particle Astrophysics, Institute of High Energy Physics, Chinese Academy of Sciences, Beijing 100049, People's Republic of China}
\correspondingauthor{Hua Feng}
\email{hfeng@ihep.ac.cn}

\author[0000-0002-4622-796X]{Roberto Soria}
\affiliation{INAF--Osservatorio Astrofisico di Torino, Strada Osservatorio 20, I-10025 Pino Torinese, Italy}
\affiliation{Sydney Institute for Astronomy, School of Physics, The University of Sydney, NSW 2006, Australia}

\author[0000-0002-1620-0897]{Fuyan Bian}
\affiliation{European Southern Observatory, Alonso de C\'ordova 3107, Casilla 19001, Vitacura, Santiago 19, Chile}

\author[0000-0001-7349-4695]{Jianfeng Wu}
\affiliation{Department of Astronomy, Xiamen University, Xiamen, Fujian 361005, People's Republic of China}

\author[0000-0002-2705-4338]{Lian Tao}
\affiliation{State Key Laboratory of Particle Astrophysics, Institute of High Energy Physics, Chinese Academy of Sciences, Beijing 100049, People's Republic of China}

\author[0000-0003-1038-9104]{Jes\'us M. Corral-Santana}
\affiliation{European Southern Observatory, Alonso de C\'ordova 3107, Casilla 19001, Vitacura, Santiago 19, Chile}

\begin{abstract}

NGC~5408 X-1 is one of the best studied ultraluminous X-ray sources (ULXs) and is surrounded by a photoionized nebula. Previous optical spectroscopy established the presence of strong Balmer, \OIII\ and \HeIIwave\ emission from the nebula, but the powering engine remains uncertain. 
In this work, we present new integral-field observations of NGC~5408 X-1, supplemented by archival long-slit spectroscopy and \textit{Hubble Space Telescope} (\textit{HST}) imaging, and confirm the presence of a composite nebula, with a small \HeIII\ region centered at the ULX and a large, shell-like \HII\ region. 
We also confirm that the broad \HeII\ emission is point-like and most likely associated with the ULX binary system. 
Photoionization simulations with \textsc{Cloudy} show that the ULX spectral energy distribution (SED), that has a total luminosity of $2.4 \times 10^{40}$\,\ergs\ and is obtained by fitting the optical/UV/X-ray data, will overpredict the \HeIII\ region luminosity and size. 
Instead, adopting the same SED shape with a reduced luminosity of $1.0\times 10^{39}$\,\ergs\ plus a blackbody with a temperature of 30000\,K and a luminosity of $1.3\times 10^{39}$\,\ergs\ can sufficiently reproduce the \HeIII\ and \HII\ regions, in terms of both luminosity and size. 
Such a dual-component ionizing spectrum is well consistent with \textit{HST} measurements of the ULX in optical and UV, while lower than the inferred X-ray luminosity assuming isotropic radiation by a factor of 24.
This implies that the EUV and X-ray emission from the ULX may be mildly beamed toward our line of sight, in line with the picture of supercritical accretion. 

\end{abstract}

\section{Introduction}
Ultraluminous X-ray sources (ULXs) are X-ray binaries with an apparent X-ray luminosity exceeding $10^{39}$~\ergs, corresponding to the Eddington limit for a $10\ M_{\odot}$ black hole \citep{Feng2011, Kaaret2017, Pinto2023, King2023}. 
Identification of pulsars in ULXs suggests that most of them are powered by supercritical accretion onto stellar mass compact objects \citep{Bachetti2014,Furst2016,Carpano2018} rather than intermediate-mass black holes.

Multi-wavelength observations are essential for clarifying the nature of supercritical accretion systems. In particular, optical observations provide an important way to study the environments of ULXs. The surrounding ionized nebulae preserve signatures of both radiative and mechanical feedback from the central source. Shock-ionized nebulae can be used to probe the mechanical impact of accretion driven outflows on the surrounding interstellar medium, while photoionized nebulae provide a means of constraining the ionizing radiation field.

Among the optical tracers, \HeIIwave\ is particularly valuable because it requires photons above 54.4~eV and therefore provides a sensitive diagnostic of hard photoionization by the ULX. For example, the dramatic variability of the \HeII\ line in NGC~1313~X-2 on timescales shorter than 24 hours indicates an origin from the outer disk, and its radial velocity may then be used to constrain the mass of the compact object under the assumption of a typical companion mass \citep{Roberts2011}. An extended \HeII\ nebula has been detected around Holmberg~II~X-1 \citep{Pakull2002}, and the \HeII\ flux enables us to place a lower limit of about $(4$--$6)\times10^{39}$~\ergs\ on the UV/X-ray luminosity of the ionizing source \citep{Kaaret2004}.

NGC~5408 X-1 is one of the best-studied ULXs with prominent \HeII\ emission \citep{Kaaret2009,Cseh2013}. The source is located in the dwarf irregular galaxy NGC~5408 at a distance of 4.8~Mpc \citep{Karachentsev2002} and reaches a peak X-ray luminosity (0.3--10~keV) of $\sim10^{40}$~\ergs\ \citep{Strohmayer2009,Hern2015}. Previous \textit{Hubble Space Telescope} (\textit{HST}) imaging identified a compact optical counterpart consistent with emission from a B0I supergiant and/or an irradiated disk \citep{Grise2012}. Optical spectroscopy revealed extended \HeII\ emission over a scale of $\sim 30$\,pc, comprising a narrow \HeIIwave\ component (FWHM $\sim75$~\kms) and a variable broad component (FWHM $\sim750$~\kms), indicating that part of the line likely originates close to the ULX system \citep{Kaaret2009,Cseh2011,Cseh2013}. 
If the line arises in the accretion disk, the observed variations put an upper limit on the semi-amplitude of the radial velocity of $K = 132\pm42$~\kms, corresponding to an upper limit of $\sim510\,M_{\odot}$ on the mass of the central compact object \citep{Cseh2011,Cseh2013}.
On larger scales, optical observations detect a shell-like nebula with a characteristic size of $\sim 60$~pc around the source \citep{Kaaret2009,Grise2012}. 
Although deep X-ray observations revealed the presence of an ultrafast outflow with a velocity of $\sim 0.2\,c$ \citep{Pinto2016}, 
emission-line ratio diagnostics suggest that the nebular environment around the ULX is predominantly photoionized rather than shock excited, \citep{Kaaret2009}.

Two questions remain: whether the full nebular complex can be explained by photoionization from the central ULX, and how the small \HeIII\ region is related to the larger shell seen in lower-ionization lines. The integral-field unit (IFU) provides spatially resolved spectroscopy and offers a direct way to address these issues. In this paper, we use new observations with the Multi Unit Spectroscopic Explorer \citep[MUSE;][]{Bacon2010} on the Very Large Telescope (VLT), together with the archival FORS2 spectroscopy and \textit{HST} imaging, to examine the morphology and ionization structure around NGC~5408 X-1. 
The paper is organized as follows. Section~\ref{sec:obs} describes the MUSE, FORS2, and \textit{HST} data and reduction procedures. 
Section~\ref{sec:nebula} presents the observational results. 
Section~\ref{sec:discuss} describes the modeling and discusses the physical implications.  
Section~\ref{sec:conc} summarizes the conclusions.

\section{Observations and Data Reductions} 
\label{sec:obs}

\subsection{VLT MUSE}

The VLT MUSE observations (PI: H.\ Feng, Program ID: 113.26LE.001) were conducted in 2024 at Paranal Observatory with two observation blocks (OBs). 
However, OB 3814176 performed on Jun 22 was not included in the analysis, because its extracted spectrum is affected by prominent spike-like residuals and a higher pixel-to-pixel noise level. This observation was obtained under nearly full-Moon conditions, with an actual fractional lunar illumination close to unity, which likely resulted in a substantially higher sky background.
OB 3814166 consisted of $4 \times 560$~s exposures, with one on May 1 and the other three on June 27. 
Between the on-source exposures, a 90-degree field rotation combined with small dithering offsets was applied.
The wide field extended mode was employed to have a wavelength coverage of 4600--9350~\AA\ and a spaxel scale of $0\farcs2 \times 0\farcs2$. 
The point-spread function measured from the reduced datacubes is about 0\farcs8 for all exposures.

For each exposure, we applied the standard {\tt EsoRex} MUSE data reduction pipeline to reduce the data. The {\tt muse\_scibasic} recipe was used to correct the bias frames, lamp flats, arc lamps, twilight flats, geometry field, and illumination exposures. Then, the {\tt muse\_scipost} was employed for telluric correction and flux calibration. Subsequent alignment and combination of individual exposures were carried out with the {\tt muse\_exp\_align} and {\tt muse\_exp\_combine} recipes, resulting in a final data cube for the OB.

We noticed that two of the June 27 exposures had systematically lower fluxes than the May 1 exposure, likely due to thin clouds during the science exposures that were not captured by the standard-star calibration. 
Using aperture photometry of several isolated point sources in line-free continuum windows, we found that the final combined cube is fainter than the May 1 reference exposure by a factor of 1.238. 
The relative scale factors show no significant wavelength dependence.
We therefore applied a multiplicative correction factor of 1.238 to the final combined cube.

\begin{figure*}
\includegraphics[width=0.33\linewidth]{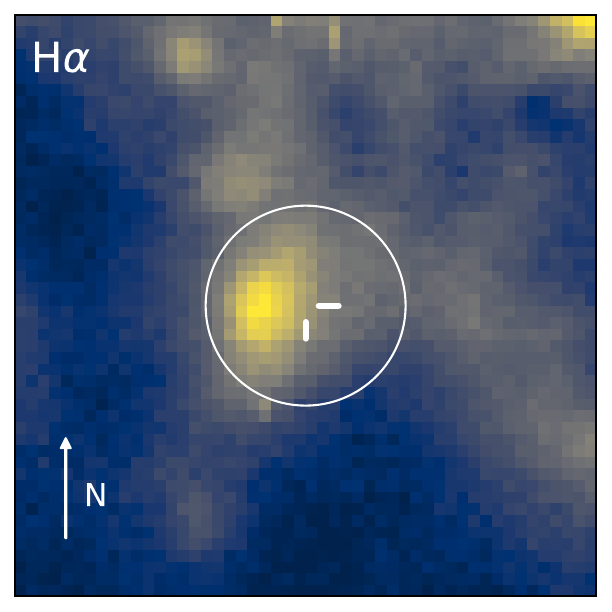}
\includegraphics[width=0.33\linewidth]{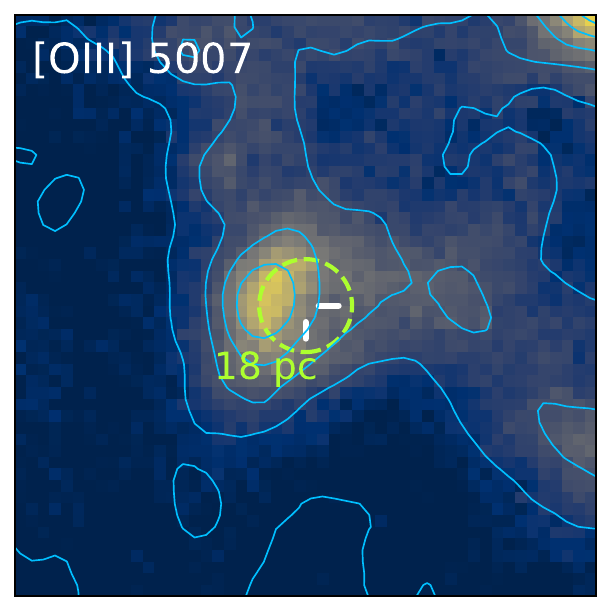}
\includegraphics[width=0.33\linewidth]{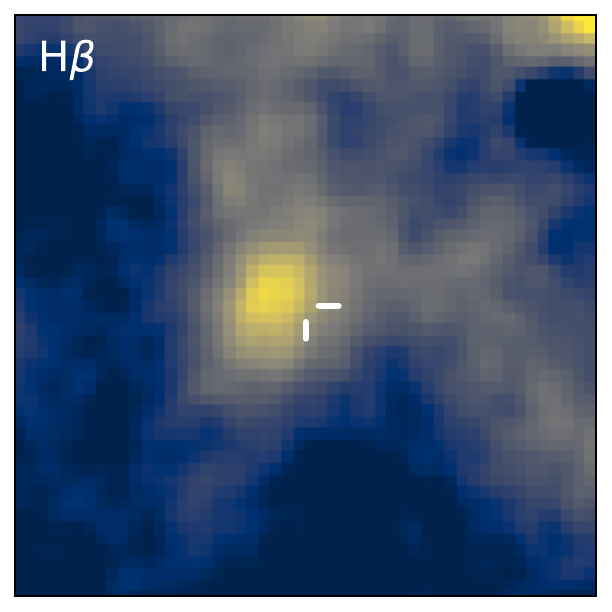} \\
\includegraphics[width=0.33\linewidth]{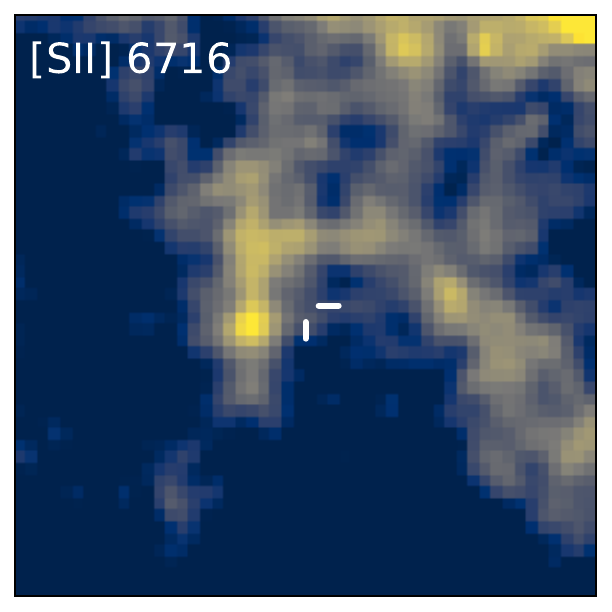}
\includegraphics[width=0.33\linewidth]{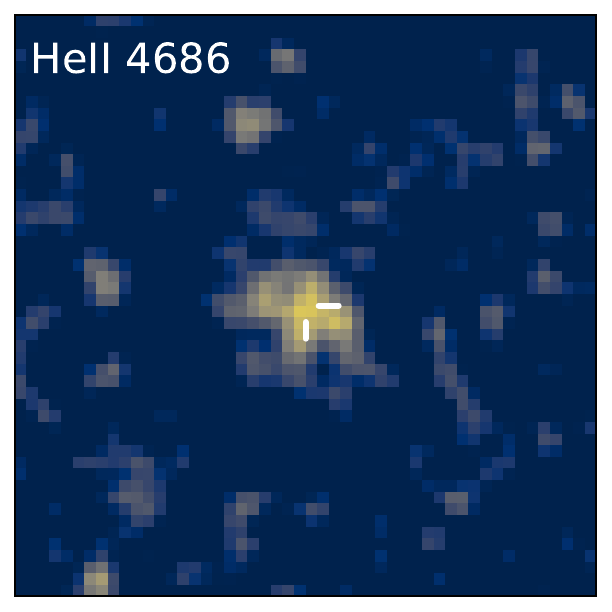}
\includegraphics[width=0.33\linewidth]{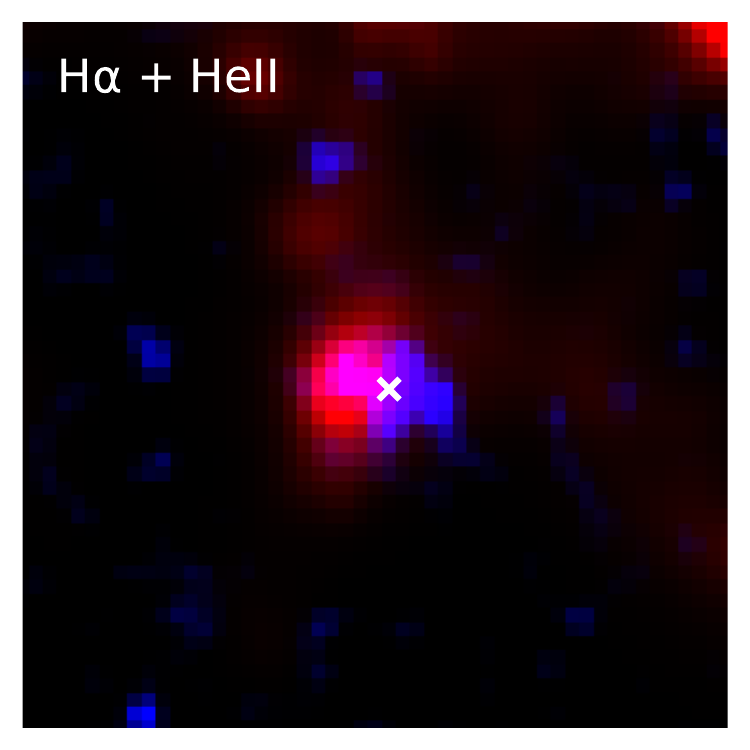}
\caption{
MUSE continuum-subtracted emission line images around NGC~5408 X-1 in \halpha, \OIIIb, \hbeta, \SIIa, and \HeII, together with a two-color image combining \halpha\ and \HeII. The white bars or cross mark the ULX position. In the H$\alpha$ panel, the white circle marks the spectral extraction aperture for the entire nebula shown in Figure~\ref{fig:spec}; the corresponding background region lies outside the displayed field. In the \OIIIb\ panel, the \halpha\  contours are overlaid in cyan, while the dashed circle with a radius of 18 pc marks the displacement of emission peak to the ULX. The \hbeta, \SIIa, and \HeII\ images are smoothed with a Gaussian kernel of $\sigma=1$ pixel. The arrow points north and has a length of 40 pc ($1\farcs72$). 
}
\label{fig:flux_maps}
\end{figure*}
 
Figure~\ref{fig:flux_maps} shows the emission line flux images, created by integrating the flux around the line centroid after subtracting the local continuum. The ULX position is determined from its optical counterpart identified in the  \textit{HST} image \citep{Grise2012} that is aligned with the MUSE images. 

\subsection{VLT FORS2}
We reanalyzed the archival VLT FORS2 spectroscopic data to investigate the \HeII\ emission at higher signal-to-noise ratio and spectral resolution. 
The observations were conducted on 2010 April 12 and consisted of three 849 s exposures. 
The instrumental setup is described in \citet{Cseh2011}. 
The data were reduced with the {\tt EsoReflex} pipeline, including bias subtraction, flat-field correction, geometric-distortion correction, wavelength calibration, and rectification to a linear wavelength scale. 

The resulting two-dimensional spectra were aligned at the ULX position and combined in Python. 
The spatial pixel scale is $0\farcs25$ pixel$^{-1}$, the spectral resolving power at \HeII ~$\lambda4686$ was $R\approx1500$, the seeing was estimated to be $\sim0\farcs5$, and the slit width was $1\farcs0$.

\subsection{\textit{HST}}

We reanalyzed the archival \textit{HST} images to better separate the compact optical counterpart from the surrounding nebular emission and obtain more accurate source fluxes. We adopted the WFC3 data (pixel size 0\farcs0396) with the F502N ([O III]$\lambda5007$; dataset IBDE04010) and F547M (dataset IBDE01010) filters, and the WFPC2 data (pixel size 0\farcs0455) with the F656N (\halpha\ and \NII; dataset UBAH2908M) and F814W filters (dataset UBAH290GN). 

\begin{figure}
\centering
\includegraphics[width=0.49\columnwidth]{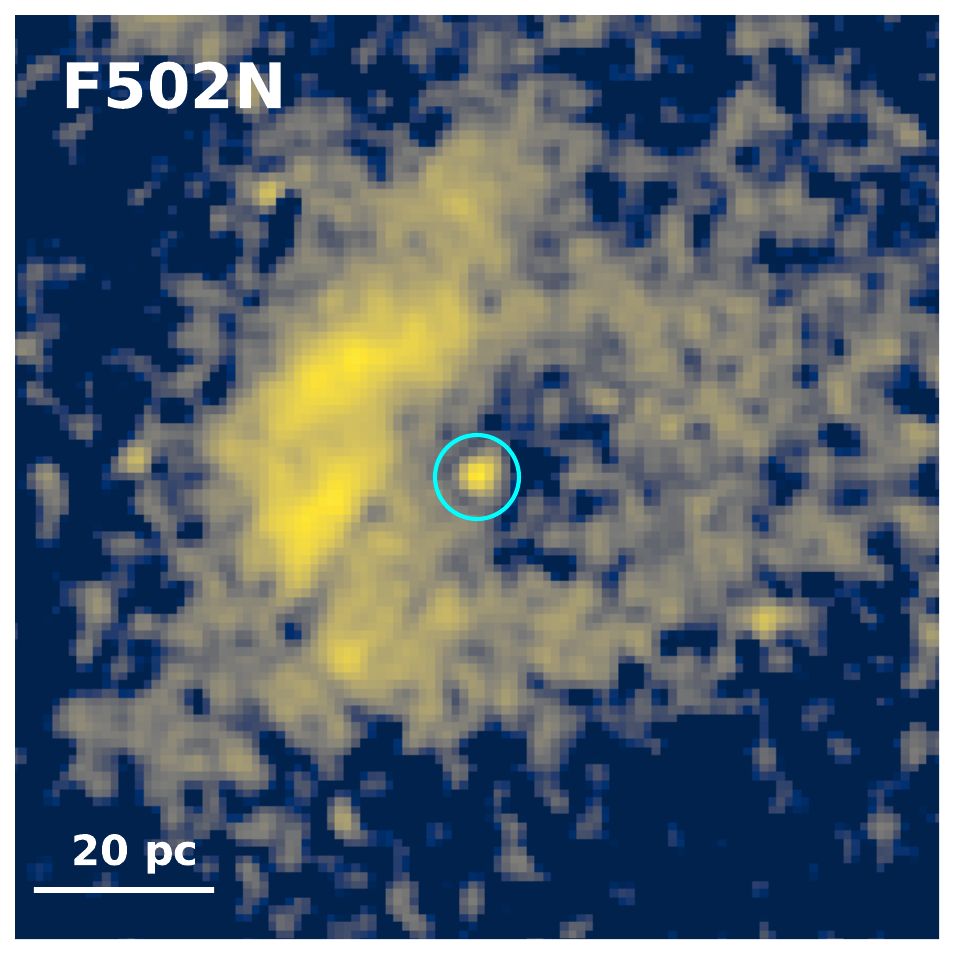}
\includegraphics[width=0.49\columnwidth]{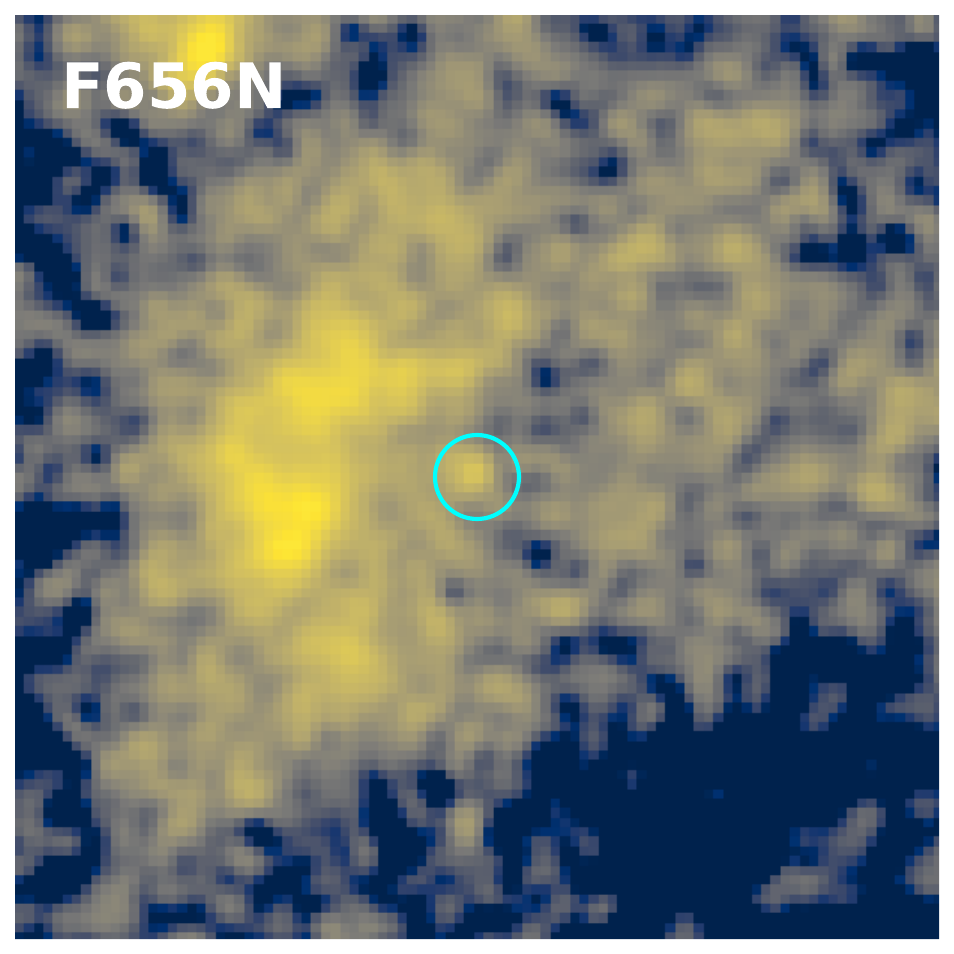}
\caption{
Continuum-subtracted \textit{HST} narrow-band images around NGC~5408 X-1, with WFC3 F502N (left) for \OIIIb\ and WFPC2 F656N (right) for \halpha\ and \NII. The images are smoothed with a Gaussian kernel of $\sigma=1$ pixel. The ULX position is enclosed by the circle.}
\label{fig:hst}
\end{figure}

We subtracted the continuum from the narrow band image by multiplying a scaling factor on the broad/medium-band image.
The scaling factor was determined by minimizing the standard deviation of point-like sources in the subtracted image, and is 0.097 for F502N/F547M and 0.017 for F656N/F814W. 
We show the continuum-subtracted images in Figure~\ref{fig:hst}. 
As one can see, the compact optical counterpart of ULX is visible in F656N and F502N, indicative of excessive \halpha\ and \OIIIb\ emission. To check whether the apparent difference between the \textit{HST}/F502N and MUSE \OIIIb\ morphologies is mainly due to spatial resolution, we convolved the \textit{HST}/F502N image with the MUSE PSF and found that the point-like \OIIIb\ source at the ULX position can no longer be seen.
The flux in the two narrow bands were measured using the {\tt Photutils} package \citep{Bradley2022} with a 2-pixel-radius aperture. 
The aperture correction is determined from isolated bright stars in the field.
We used {\tt Synphot} to derive the emission line flux given a Gaussian line profile. 
The fluxes of the ULX counterpart measured from the continuum-subtracted images are $(1.3\pm0.8)\times10^{-16}$~\ergcms\ in F656N and $(1.52\pm0.04)\times10^{-16}$~\ergcms\ in F502N. 
Because the F656N image exhibits an irregular point-spread function (PSF), reliable PSF photometry could not be performed. 
Aperture photometry is also subject to relatively large uncertainties owing to the complex environment around X-1. 
Nevertheless, these measurements indicate that the \halpha\ and \OIIIb\ emission from the ULX counterpart is relatively weak compared with that of the surrounding nebula.

\section{The Composite Nebula}
\label{sec:nebula}

Figure~\ref{fig:flux_maps} presents zoomed-in images around NGC 5408 X-1 in \halpha, \OIIIb, \hbeta, \SIIa, and \HeII. The \halpha, \OIIIb, \hbeta, and \SIIa\ images reveal a shell-like nebula with a characteristic size of $\sim$60~pc.
The peak position is offset from the ULX, which has previously been noticed in the long-slit spectrum \citep{Kaaret2009}.
In contrast, the \HeII\ emission is spatially coincident with the ULX position.

\begin{figure}
\centering
\includegraphics[width=1.\columnwidth]{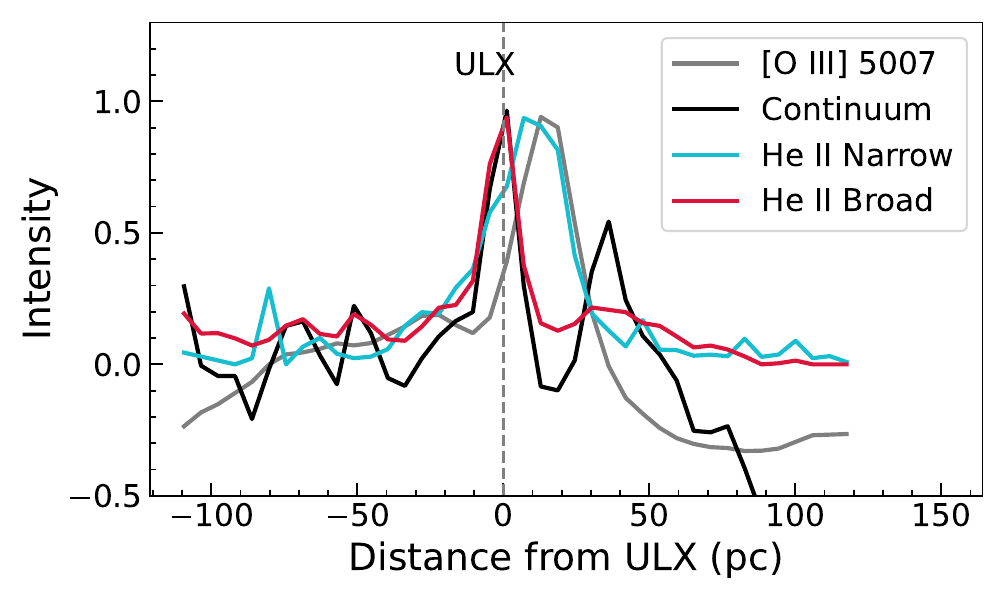}
\caption{Spatial profiles of emission lines and the continuum derived from the FORS2 data, with the peak normalized to unity. The continuum component in the emission lines has been subtracted. 
Note that the second peak in the continuum near 35 pc is due to a point-like star.
The vertical dashed line marks the position of ULX. 
}
\label{fig:profile_fors2}
\end{figure}

We further constructed spatial profiles of \OIIIb, the adjacent continuum, and the narrow and broad \HeII\ components from the FORS2 data, as shown in Figure~\ref{fig:profile_fors2}. 
The continuum contribution was subtracted from the emission-line profiles. 
For the \HeII\ emission, we decomposed the line into narrow and broad components to examine their spatial distributions separately. 
We first fitted the spectra extracted from the three spatial pixels with the highest signal-to-noise ratios using a two-component Gaussian model. 
We then fixed the best-fit centroids and widths of both components and fitted the spectrum from each spatial pixel, allowing only the normalizations to vary.

As shown in Figure~\ref{fig:profile_fors2}, the spatial profile of the broad component is consistent with that of the continuum, indicating that the broad component is point-like and likely originates from the ULX binary system. 
In contrast, the narrow component is spatially extended. 
After correcting for the seeing FWHM, a Gaussian fit to the spatial profile of the narrow component gives an intrinsic FWHM of $\sim 28$\,pc, consistent with the FORS1 result (30 pc) reported by \citet{Kaaret2009}.

The FORS2 slit has a width of $1\farcs0$, which does not fully cover the entire \HeII-emitting region with an observed size of $\sim1\farcs3$. Assuming that the \HeII-emitting region is approximately circular and centered on the slit, the slit covers about 88\% of its projected area. Applying this approximate geometrical correction, we estimate the total narrow-component \HeII\ flux from the extended region to be $(3.13\pm0.33)\times10^{-16}$~\ergcms. 
For the broad component, which is point-like, we directly measured a flux of $(1.24 \pm 0.45) \times 10^{-16}$\,\ergcms.

For comparison, we also estimated the total \HeII\ flux from the MUSE data using an aperture that 
fully encloses the \HeII-emitting region. Because the MUSE data cannot spectrally resolve the narrow and broad \HeII\ components, the measured value corresponds to the total \HeII\ flux, $(4.30\pm0.36)\times10^{-16}$~\ergcms. Given the lower S/N of the MUSE data, this value is consistent with the total FORS2 \HeII\ flux. In the following analysis, we therefore adopt the narrow-component FORS2 measurement, $(3.13\pm0.33)\times10^{-16}$~\ergcms, as the flux of the extended, narrow \HeII-emitting region.

\begin{figure*}
\centering
\includegraphics[width=2\columnwidth]{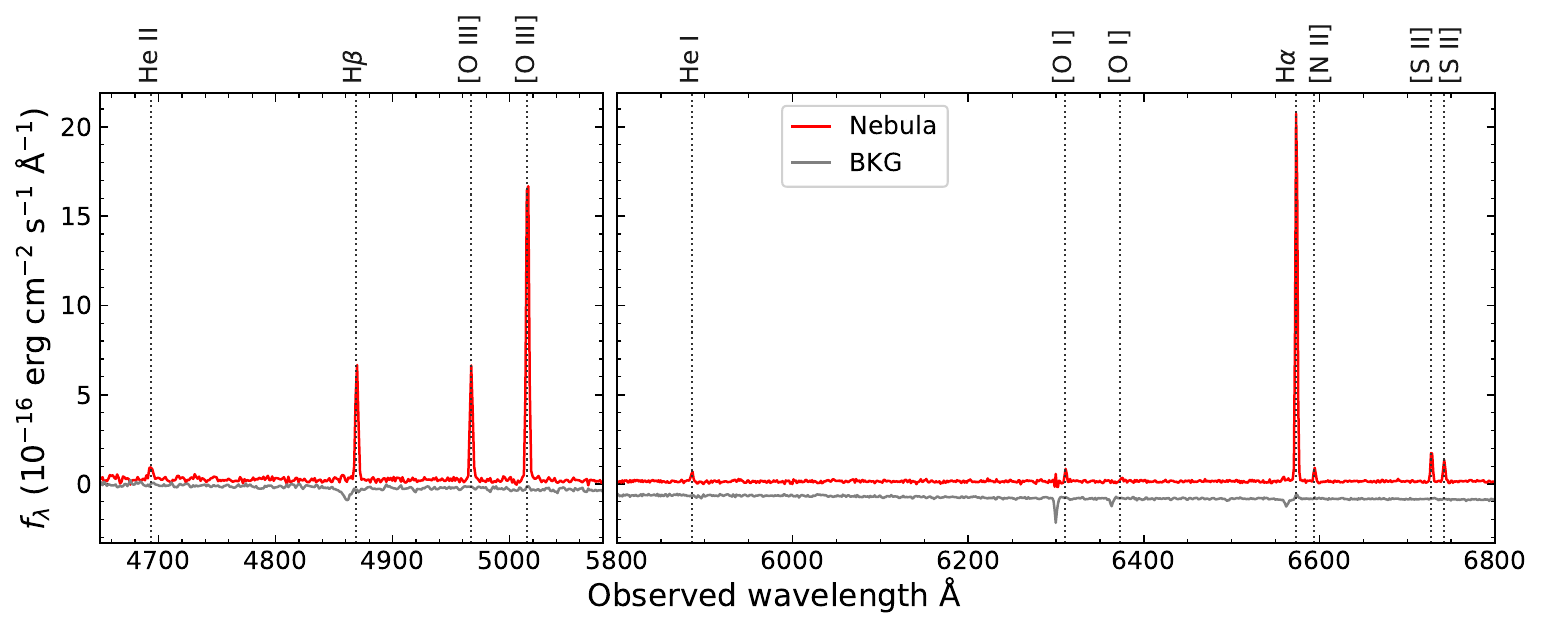}
\caption{MUSE spectra of the whole nebula. The background spectrum (gray) is shifted for display.}
\label{fig:spec}
\end{figure*}

We measured the emission-line fluxes from the MUSE, FORS2, and \textit{HST} observations. 
For the MUSE data, the spectrum of the whole nebula was extracted using a circular aperture with a radius of $1\farcs72$, corresponding to 40 pc, centered on the ULX position and encompassing the entire nebula. 
The background was estimated from a nearby source-free region. 
The extraction aperture is indicated in the \halpha\ panel of Figure~\ref{fig:flux_maps}. 
The extracted spectrum is shown in Figure~\ref{fig:spec}, and the measured line fluxes are listed in Table~\ref{tab:linesflux}. 
Because the MUSE data do not spatially resolve the ULX counterpart from the surrounding nebula, the measured fluxes may include a contribution of 2\% from the counterpart estimated from the \textit{HST} observations. 

\begin{deluxetable}{lcc}[tb]
\tablecaption{Observed emission-line fluxes and predictions from the fiducial dual-component photoionization model.}
\tablewidth{\columnwidth}
\label{tab:linesflux}
\tablehead{
\colhead{Emission line} &
\colhead{Observed} &
\colhead{Model predicted}
}
\startdata
\halpha\ & $79.74\pm0.18$ & 82.78 \\
\OIIIb\ & $78.18\pm0.28$ & 65.49 \\
\hbeta\ & $26.55\pm0.27$ & 28.87 \\
\HeII\ (narrow) & $3.13\pm0.33$ & 3.57 \\
\enddata
\tablecomments{Fluxes are in units of $10^{-16}$\,\ergcms. 
The \halpha, \hbeta, and \OIIIb\ fluxes are measured from the MUSE data, while the narrow-component \HeII\ flux is derived from the FORS2 spectrum, all corrected for interstellar reddening assuming $E(B-V)=0.08\pm0.03$ \citep{Cseh2011}, using the extinction law of \citet{Cardelli1989} with $R_V=3.1$. 
}
\end{deluxetable}

\section{Discussion}
\label{sec:discuss}

With the MUSE observations as well as the archival FORS2 and \textit{HST} data, we identified a composite nebula including two regions around NGC 5408 X-1, a large ($\sim$ 2$\farcs$6, 60 pc in diameter) shell-like \HII\ region with narrow (FWHM $\approx$ 60\,\kms) Balmer and \OIII\ emission lines, and a small ($\sim$ 1$\farcs$3, 30 pc in diameter) \HeIII\ region with narrow (FWHM $\approx$ 75\,\kms) \HeII\ emission centered at the ULX position. The broad \HeII\ emission (FWHM $\approx$ 750\,\kms) is point-like and most likely associated with the binary system, and will not be discussed in the following. 

First of all, we argue that shock ionization cannot be the major powering mechanism for the composite nebula. 
The shock velocity is related to the emission line width as 0.425\,FWHM (lower limit) in the case of thin-shell radiative shock, or 0.57\,FWHM (upper limit) for adiabatic shock \citep[see][]{Soria2021}. 
This suggests that the shock velocity is less than 42.75\,\kms\ for the small \HeIII\ region or less than 34.2\,\kms\ for the large one. 
Such low velocities are unable to power high excitation lines such as \HeII\ or \OIII\ \citep{Allen2008}.

We therefore performed photoionization modeling with \texttt{pyCloudy} \citep{Morisset2013}, a Python interface to the spectral synthesis code \texttt{Cloudy} \citep{Ferland2017}. 
The electron density was estimated from the \SIIa/\SIIb\ flux ratio
$1.51\pm0.07$ using PyNeb \citep{Luridiana2015}. The measured ratio is consistent with the low-density limit of the \SII\ diagnostic, implying $n_{\rm e}\lesssim20\ {\rm cm}^{-3}$.
Guided by this constraint, we explored a hydrogen density grid from $n_{\rm H}=1$ to $20\ {\rm cm}^{-3}$ in steps of $1\ {\rm cm}^{-3}$.
We also tested models with $n_{\rm H}>20\ {\rm cm}^{-3}$ and found that they produce a \HeIII\ region much smaller than the observed $\sim30$\,pc scale.

We assumed a spherical geometry with the ULX located at the center. The inner radius of the surrounding gas was set to $\sim1$ pc, while the outer radius was set to $50$ pc in order to encompass the entire nebula, including the central \HeIII\ region and the more extended \HII\ region. 
We adopted a metallicity of $0.07\,Z_{\odot}$, following \citet{Soria2004}. 
We also independently estimated the oxygen abundance using several empirical strong-line diagnostics. These methods indicate a low-metallicity environment, with $12+\log({\rm O/H})\approx7.9$--8.1, although the exact abundance remains uncertain because the nebula is photoionized by a ULX system rather than by normal massive stars, for which most empirical calibrations were established.

We therefore used $Z=0.07\,Z_\odot$ as the fiducial metallicity in our \texttt{Cloudy} calculations and further explored models with higher metallicities. The model predictions are relatively insensitive to metallicity over $0.07$--$0.10\,Z_\odot$, whereas models with $Z\gtrsim0.1\,Z_\odot$ increasingly overpredict the observed \OIIIb\ emission. In particular, for $Z=0.2\,Z_\odot$, the best model that is consistent with the observed morphology predicts an \OIIIb\ flux $\sim1.7$ times that of \halpha, larger than observed.

\begin{figure}[tb]
\centering
\includegraphics[width=0.9\columnwidth]{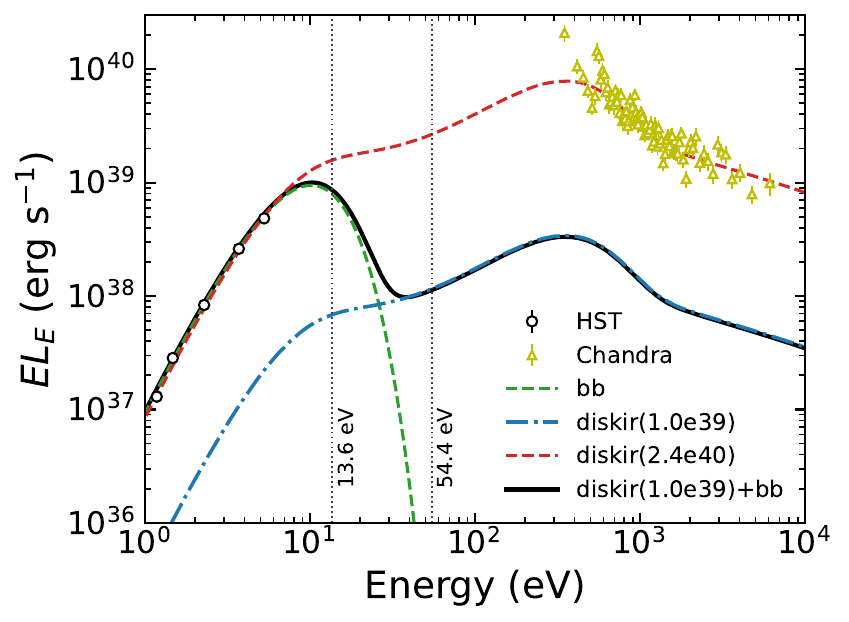}
\caption{Ionizing spectra adopted in the \textsc{Cloudy} calculations: the original {\tt diskir} SED with a total luminosity of $2.4 \times 10^{40}$\,\ergs, a scaled-down {\tt diskir} SED with a total luminosity of $1.0 \times 10^{39}$\,\ergs\ plus a blackbody component at 30000\,K.
Data are adopted from \cite{Grise2012}, including dereddened \textit{HST} photometric measurements shown as open circles, and unabsorbed \textit{Chandra} spectrum in open triangles.
The vertical dotted lines indicate the ionization energies of H\,I (13.6 eV) and He\,II (54.4 eV).
}
\label{fig:sed}
\end{figure}

\begin{figure}
\centering
\includegraphics[width=0.9\columnwidth]{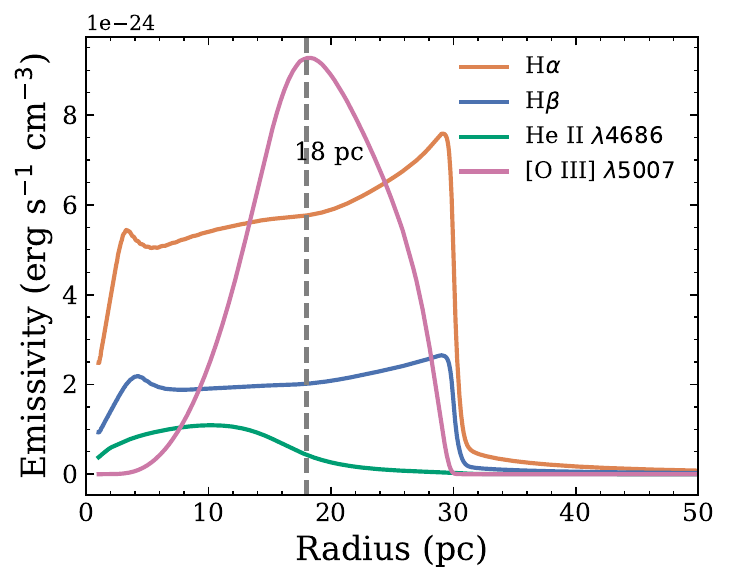}
\caption{
Radial emissivity profiles predicted by \textsc{Cloudy} with the dual-component model. 
The dashed line marks the peak \OIIIb\ emissivity position ($\sim18$\,pc), matching the observed peak displacement (see Figure~\ref{fig:flux_maps}).
}
\label{fig:cloudy}
\end{figure}

For the ionizing spectrum, we first adopted the {\tt diskir} SED presented by \citet{Grise2012}. The model has $\Gamma=2.42$, $kT_{\rm in}=0.13$~keV, $kT_{\rm e}=50$~keV, $L_{\rm C}/L_{\rm D}=0.57$, and $f_{\rm out}=0.043$. It was fitted to the optical and X-ray observations and is shown as the red curve in Figure~\ref{fig:sed}, with a total luminosity of $\sim2.4\times10^{40}$\ \ergs. However, such an ionizing source predicts \HeII\ emission extending over a size of $\sim$60 pc, whereas the observed \HeIII\ region has a diameter of only $\sim$30 pc, and also overpredicts the \HeII\ luminosity by a factor of $\sim$10.

To fit the \HeIII\ region, we kept the SED shape but scaled down the luminosity, and found that a total luminosity of $1.0\times 10^{39}$\,\ergs\ can just match the observations. The new SED is shown as the blue curve in Figure~\ref{fig:sed}. 
In this case, models with $n_{\rm H}\approx4$--$10\ {\rm cm}^{-3}$ can reproduce the observed \HeII\ flux and size, consistent with the density estimate of \citet{Kaaret2009}.
However, the scaled-down {\tt diskir} model alone cannot reproduce the extended Balmer and \OIII\ emission. With the density $n_{\rm H}\lesssim4\ {\rm cm}^{-3}$, the \HII\ region can extend to $\sim60$ pc, comparable to the observed size; however, the predicted \halpha\ and \OIIIb\ fluxes reach only $\sim50\%$ and $\sim20\%$ of the observed values, respectively. This deficit suggests that an additional ionizing source is required.

We therefore added a second, blackbody ionizing source in addition to the scaled-down {\tt diskir} component. We explored a parameter grid in which the blackbody temperature varies from $T_{\rm bb}=20000$ to $45000$ K in steps of $2500$ K, the ionizing photon rate varies from $\log [Q({\rm H})/({\rm ph\ s^{-1}})]=40$ to $50$ in steps of $0.5$ dex, and the gas density varies from $n_{\rm H}=4$ to $10\ {\rm cm}^{-3}$ in steps of $1\ {\rm cm}^{-3}$. We found that the blackbody component contributes little to the central \HeIII\ region, but significantly affects the larger \HII\ region by providing additional photons between 13.6 eV and 54.4 eV. Models with $n_{\rm H}\sim5\ {\rm cm}^{-3}$ and $\log [Q({\rm H})/({\rm ph\ s^{-1}})] = 49$ provide the best agreement with the observations in terms of both line fluxes and emission size. With $\log [Q({\rm H})/({\rm ph\ s^{-1}})] = 49$, the predicted line ratios are relatively insensitive to the blackbody temperature over $T_{\rm bb}=20000 - 45000$ K. The optical/UV continuum of the ULX counterpart measured with \textit{HST} therefore provides an additional constraint: it can be described by a blackbody with $T_{\rm bb}=30000$\,K and $L_{\rm bb}=1.3\times10^{39}$\,\ergs, as shown in Figure~\ref{fig:sed}. This temperature is also consistent with the interpretation by \citet{Grise2012}, who found that the optical/UV emission is compatible with a B0\,I supergiant with an effective temperature of $\sim30000$\,K.

We therefore adopted this model as a fiducial model, i.e., {\tt diskir} with a total luminosity of $1.0 \times 10^{39}$\,\ergs plus a blackbody with $T_{\rm bb} = 30000$\,K and $L_{\rm bb} = 1.3 \times 10^{39}$\,\ergs, and show its predicted radial emissivity profiles in Figure~\ref{fig:cloudy}. In this model, the \HeIII\ region is confined to the inner $\sim30$ pc in diameter, consistent with observations. In contrast, the Balmer-emitting region extends over a larger scale, reaching a diameter of $\sim60$ pc. The \OIIIb\ emissivity peaks at a radius of $\sim18$ pc, well consistent with the observed morphology shown in Figure~\ref{fig:flux_maps}. The predicted fluxes from this representative model are compared with the observed values in Table~\ref{tab:linesflux}. The model reproduces the observed fluxes of \halpha, \hbeta, \OIIIb, and \HeII\ to within $\sim5$--$15\%$.

The Cloudy simulations also reveal that the radial distribution of \OIIIb\ is narrower than that of \halpha. 
The \OIIIb-to-\halpha\ ratio is moderately above unity at the peak position but slightly less than unity (around 0.9) as a whole.
This is in reasonable agreement with the spatial distribution seen in the \textit{HST} images and the total flux quoted in Table~\ref{tab:linesflux}.

Our simulations extend the photoionization analysis of \citet{Kaaret2009}. Similar to their study, we find that the observed line emission requires an ionizing luminosity substantially lower than the observed line-of-sight X-ray luminosity. However, while a reduced ULX luminosity is sufficient to reproduce the observed \HeII\ emission, it fails to account for the luminosities and spatial extents of the more extended \halpha\ and \OIII\ nebula. By introducing an additional stellar blackbody component motivated by the HST observations, our model simultaneously reproduces the observed \HeII\ luminosity and size together with the fluxes and spatial distributions of the lower-ionization nebular emission.

In addition to the two-component model, we further investigate if the original {\tt diskir} model with extra EUV absorption can provide sufficient ionization. 
The model must explain the observed optical/UV and X-ray spectra.
We therefore constructed a modified SED in which the X-ray (> 0.3 keV) and optical/UV (< 10 eV) bands follow the original {\tt diskir}, whereas the 10--150 eV portion is replaced by the fiducial model. The intervening 150--300 eV range was connected by a straight line in logarithmic space between the two components, ensuring a continuous SED.
This construction was intended to reduce the EUV photons (above 54.4 eV) but maintain the UV photons (above 13.6 eV) such that both the \HeIII\ and \HII\ regions can be explained. 
However, no matter how we fine-tune the modification (on the basis that the model still explains the data), we cannot find a solution to explain both regions.

As extra EUV absorption cannot explain the observations, a natural explanation is that the discrepancy between the line-of-sight ULX luminosity and the lower ionizing luminosity required by the nebula is due to anisotropic emission of EUV and X-ray from the ULX.  
This aligns with the standard picture that, under supercritical accretion, a central funnel is formed and emission from the central accretion flow is geometrically beamed along the symmetric axis \citep[see the review of][and references therein]{Kaaret2017}. 
The detection of ultrafast outflows in NGC 5408 X-1 \citep{Pinto2016} is consistent with the presence of a supercritical accretion flow and such a disk-wind geometry.
In this scenario, the luminosity seen by the surrounding gas is about 1/24 of the line-of-sight luminosity under the isotropic assumption. Otherwise, if the ULX emission were indeed isotropic, the discrepancy would require a nebular covering factor of $\lesssim$1/24. Such a small value could arise from a thin disk, but appears less likely given the shell-like morphology.

We note that we have assumed a spherical geometry in the \textsc{Cloudy} calculations, while the HST narrow-band images show that the shell-like nebula is broadly round but asymmetric, with the east side brighter than the west. 
We found that, lower gas densities (e.g., $n_{\rm H}\sim3\ {\rm cm^{-3}}$) shift the \OIIIb\ emissivity peak to larger radii and reduce the surface brightness, and vice versa. Thus, the observed surface-brightness asymmetry may arise from variations in the surrounding gas density, with the fainter side possibly having a lower density.

An alternative possibility is that the observed X-ray spectrum does not represent the long-term average ionizing continuum responsible for powering the nebula. 
The recombination timescale is mainly scaled with the electron density as $\tau_{\rm rec} = 1.2 \times 10^5 / (n_{\rm e} / {\rm cm^{-3}})$\,yr, assuming a constant recombination coefficient which is weakly related to the temperature. We thus estimate a recombination timescale of $\sim 2 \times 10^{4}$\,yr given $n_{\rm e} \simeq 5\; {\rm cm^{-3}}$, significantly longer than typical observation timescales (tens of years).
In this scenario, the ULX may have undergone on/off switches and is actively only recently.

\section{Conclusions}
\label{sec:conc}

We summarize our main results as follows.

\begin{itemize}[nosep]

\item Surrounding NGC~5408 X-1, we confirmed the presence of a composite nebula consisting of a large, shell-like \HII\ region with an extent of $\sim60$ pc, traced by prominent \halpha, \hbeta, and \OIII\ emission (FWHM $\approx 60$\,\kms), and a smaller \HeIII\ region (FWHM $\approx 75$\,\kms) with an extent of $\sim30$ pc centered at the ULX position. 

\item The broad \HeII\ component (FWHM $\approx 750$\,\kms) is point-like and coincident with the ULX position, most likely originating from the ULX binary system. 

\item The narrow line widths, together with the presence of high-excitation lines such as \HeII\ and \OIII, suggest that shocks are unlikely to be the dominant ionization mechanism of the nebula.

\item Photoionization from the ULX SED with a total luminosity of $2.4 \times 10^{40}$\,\ergs\ overpredicts the \HeIII\ region size and luminosity. Scaling down to $1.0\times 10^{39}$\,\ergs, plus a blackbody component with $T_{\rm bb} = 30000$\,K and $\log [Q({\rm H})/({\rm ph\ s^{-1}})]=49$ ($L_{\rm bb} = 1.3\times10^{39}$\,\ergs) can reproduce the observed sizes and luminosities of both regions. The scaled-down ULX SED is mainly responsible for the central \HeIII\ region while the blackbody mainly for the outer \HII\ region. 

\item The dual-component ionizing spectrum fits the observed optical/UV data but is 24 times lower than the observed X-ray spectrum, suggesting that the EUV and X-ray emission from the ULX may be mildly beamed. 

\end{itemize}

\end{CJK*}
\bibliography{NGC5408}{}
\bibliographystyle{aasjournal}

\end{document}